\documentclass[aps,prb,showpacs,twocolumn,groupedaddress]{revtex4}
\usepackage[english]{babel}
\usepackage{graphicx}
\usepackage{color}
\usepackage[utf8]{inputenc}
\usepackage{pstricks,pst-grad,color}
\usepackage{graphicx,pifont,amssymb}
\usepackage{amsmath,amssymb}

\def\la{\langle}
\def\ra{\rangle}

\begin{document}

\title{Ferromagnetic state in the one-dimensional Kondo lattice model} 



\author{Robert Peters}
\email[]{peters@scphys.kyoto-u.ac.jp}
\affiliation{Department of Physics, Kyoto University, Kyoto 606-8502, Japan}
\author{Norio Kawakami}
\affiliation{Department of Physics, Kyoto University, Kyoto 606-8502, Japan}

\date{\today}


\begin{abstract}
In our recent study, Phys. Rev. Lett. {\bf 108} 086402 (2012), we have
revealed the intriguing properties of the ferromagnetic state
 in the Kondo lattice model 
with antiferromagnetic coupling in infinite dimensions:  
within the ferromagnetic metallic phase, the minority conduction electrons
form a gap at the Fermi energy and do not participate in transport irrespective of interaction strength and filling. This half-metallic state is 
referred to as a spin-selective Kondo insulator.  We here show that the spin-selective 
Kondo insulator can also be realized
in the one-dimensional Kondo lattice model by studying static and dynamical 
quantities with the density matrix renormalization
group. The emergence of the spin selective Kondo insulator both 
in one and infinite dimensions certainly demonstrates that this mechanism is 
quite general and ubiquitous for the ferromagnetic state in the
 Kondo lattice model irrespective of the dimensionality of the system.
\end{abstract}


\maketitle

\section{Introduction}
The Kondo lattice model is one of the fundamental models for strongly
correlated materials. 
It describes a lattice of local moments
which are coupled locally to itinerant electrons. The Hamiltonian can
be written as\cite{doniach77,lacroix1979,fazekas1991} 
\begin{displaymath}
H=t\sum_{<i,j>\sigma}c^\dagger_{i\sigma}c_{j\sigma}+J\sum_i\vec{S}_i\cdot\vec{s}_i,
\end{displaymath}
where the first term describes electron hopping on a lattice and the
second term the spin-spin interaction between a localized moment
$\vec{S}$ and the spin of a conduction electron,
$\vec{s}_i=c_{i\rho}^\dagger\vec{\sigma}_{\rho\rho^\prime}c_{i\rho^\prime}$
(with Pauli matrices $\vec{\sigma}$).

The Kondo lattice model is widely used for describing strongly
correlated electron systems.
Examples range from transition metal oxides such as the 
manganites\cite{Dagotto2001} to heavy fermions.
In the case of the manganites, the
{\it d}-orbitals are split due to crystal fields into $e_g$- and
$t_{2g}$-orbitals. Because of strong 
local interactions the half-filled $t_{2g}$-orbitals, which lie
energetically lower than the $e_g$-orbitals, can be
approximated as localized spins,
 which are ferromagnetically coupled to the electrons in the
 $e_g$-band due to Hund's rule.  
Other examples are materials with partially filled
{\it f}-orbitals, in which the electrons
can be usually very well approximated as localized moments. However, 
a local hybridization between the {\it f}-orbitals and the
itinerant electrons gives rise to an effective antiferromagnetic 
interaction between them at low
temperatures. This second example describes systems such as
heavy fermions and Kondo insulators. This article concentrates on
the latter materials, which can be described by an
antiferromagnetically coupled Kondo lattice model.

In our recent study,\cite{peters2012} we have uncovered the intriguing
properties of the ferromagnetic state in the Kondo lattice model
(with antiferromagnetically coupled spins), which
we call the ``spin-selective Kondo insulator''. 
It has been shown
that while the majority conduction electrons are metallic, the
minority conduction electrons form a gap at the Fermi energy, thus
being insulating at low temperatures. 
The existence of such half-metallic states in the Kondo lattice
model has been proposed in earlier
works.\cite{Irkhin1991,Beach2008,Kusminskiy2008} 
In our recent study we have demonstrated that the gap width follows a
Kondo-type functional form, $T_K\sim \exp(-1/J)$.
Furthermore, we have clarified that the origin of this remarkable state
is the cooperation between the localized spins and the conduction
electrons. Namely, they adjust themselves in a way that the sum of the
minority-spin conduction electrons and the localized-electrons with the 
same spin always satisfies the condition 
of ``one electron per lattice site''. Remarkably, this commensurate
situation is not only generated for a particular filling or 
combination of interaction parameters, but is ubiquitous for the
ferromagnetic state in the antiferromagnetically coupled Kondo lattice
model. 

The above conclusion  was drawn,\cite{peters2012} however,  via the
dynamical mean field theory (DMFT).\cite{georges1996} The DMFT method,
which maps the lattice model onto an impurity model embedded in a bath
to be solved self-consistently, completely neglects spatial
fluctuations, becoming exact in the limit of infinite
dimensions. Therefore some serious questions naturally arise; whether
the above conclusion depends on the method employed (infinite
dimensions); what is really expected for low-dimensional systems, etc.

In this article, we address these questions by considering the opposite extreme, {\it i.e.} the one-dimensional (1D) Kondo lattice model,
 and demonstrate that the notion of the spin-selective Kondo insulator is fundamental for Kondo lattice systems. 
We explicitly show the emergence of the 
spin selective Kondo insulator in the ferromagnetic phase 
of the 1D lattice model by using 
the density matrix renormalization group
(DMRG),\cite{white1992,schollwoeck2005,schollwoeck2011}
 which provides an excellent 
tool to calculate the ground state quantities in 1D with very high precision.
An important point is that DMRG can precisely take into account spatial fluctuations that were neglected in the DMFT study.

The paper is organized as follows. After revisiting the ground state phase 
diagram in 1D briefly in Sec.II, we analyze static as well as dynamical quantities of the 1D Kondo lattice model in Sec. III. We also compare the present 
results in 1D with those obtained by DMFT, and show that the ferromagnetic state possesses very similar properties for both cases.
Through these comprehensive analyses in 1D, together with 
the previous DMFT results in infinite dimensions, we demonstrate that
the spin-selective Kondo insulator is  general and fundamental 
for the ferromagnetic state 
of the Kondo lattice model in any dimension. 
Sec. IV is devoted to a brief summary of the paper.

\section{Phase diagram at zero temperature revisited}

\begin{figure}[tb]
\includegraphics[width=\linewidth]{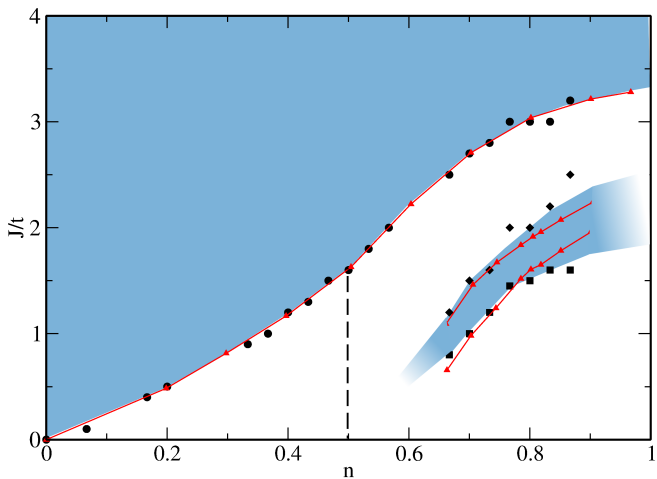}
\caption{(Color online) Phase diagram of the 1D Kondo
  lattice model. Black symbols represent the phase boundaries which are
  obtained in this work. The calculations are performed
  using $L=60$- and 
  $L=120$-site chains and $m=800-1200$ states for the density matrix. Only
  Abelian  symmetries are taken into account.
  Red lines and symbols are taken from 
  \textcite{mcculloch2002} Blue (white) color represents the
  ferromagnetic (paramagnetic) phase.
 \label{oneD_phase}}
\end{figure}

Let us start with an overview of the magnetic phase diagram of
the 1D Kondo lattice model away from half filling. 
We assume that
the coupling between localized spins and conduction electrons
is antiferromagnetic. 
The phase diagram for the 1D Kondo
lattice model has been calculated by a number of groups
before.\cite{troyer1993,tsunetsugu1997,mcculloch2001,mcculloch2002,gulasci2004,basylko2008} 
In order to clarify the parameter regime where the
ferromagnetism is stabilized, we show the phase diagram 
in Fig. \ref{oneD_phase}.
 We compare our phase boundaries to those of \textcite{mcculloch2002}
 obtained by using non-Abelian DMRG with SU(2) symmetry (note that our 
results resort to only Abelian symmetries). 
The phase diagram includes two ferromagnetic phases away from half
filling. The first 
ferromagnetic phase extends from weak coupling and low filling of
the conduction electrons. For strong coupling, this ferromagnetic phase
extends in the whole metallic region away from half filling. For intermediate
coupling within the paramagnetic phase, there is another
ferromagnetic phase, which was firstly identified by using non-Abelian
DMRG.\cite{mcculloch2002}
It was stated that this phase (intermediate coupling strengths and
fillings $n_c>0.5$) is very difficult to observe in the DMRG
calculations conserving only
Abelian symmetries.\cite{mcculloch2002} In the present study, however, we 
can clearly observe this ferromagnetic phase at intermediate coupling
strengths. 

\begin{figure}[tb]
\includegraphics[width=\linewidth]{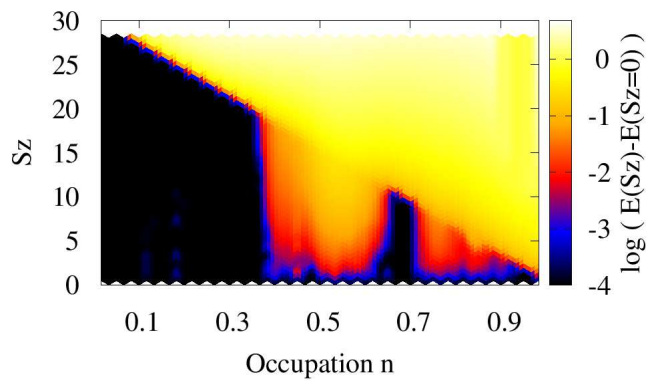}
\caption{(Color online) Energy differences on a logarithmic scale for
  the 1D 
  Kondo lattice model calculated with DMRG for $J=t$ and $m=800$ states kept
  within DMRG. For each
  filling the energy for $S_z^{tot}=0$ ($E(S_z^{tot}=0)$) is subtracted. In this way
  the ferromagnetic phases can be clearly identified without
  using SU(2)-symmetry-conserving calculations.
 \label{DMRG_energ}}
\end{figure}

Our strategy is as follows.  To identify the
ferromagnetic phases we first determine the energetically lowest eigenstate 
of all $S^{tot}_z$-sectors, and calculate the energy difference between
different 
$S_z^{tot}$-sectors, $E(S_z^{tot}\neq 0)-E(S_z^{tot}=0)$. For a
ferromagnetic state, all states fulfilling $S_z^{tot}\leq \mathcal{S}^{tot}$ 
($\mathcal{S}^{tot}$ is the SU(2) spin quantum number of the ground state) 
are supposed to be  energetically degenerate.
Plotting these energy differences for different fillings on a
logarithmic scale, one 
can clearly identify two distinct ferromagnetic phases, where the energies
do not change for $S_z^{tot}\leq \mathcal{S}^{tot}$. 
An example for this procedure for
$J=t$ is shown in Fig. \ref{DMRG_energ}.

In this way, we can deduce the phase diagram of the 1D Kondo
lattice model containing two ferromagnetic phases (blue area in
Fig. \ref{oneD_phase}). Having found good agreement in the phase
boundaries of the ferromagnetic phases with
those obtained by the SU(2) calculations,\cite{mcculloch2002} we 
believe that our calculations are accurate enough to analyze the detailed 
structure of the ferromagnetic states.

\section{Ferromagnetic state at low fillings}
\subsection{static quantities}
In the following we focus on the ferromagnetic state in the
1D Kondo lattice model. 
In the strongly coupled Kondo lattice 
with $ J\rightarrow \infty$, all electrons
are supposed to be bound in singlet states.
Nevertheless, in order to create a ferromagnetic state, a
fraction of the electrons should not be bound in singlets. 
These electrons ``connect'' the different lattice sites and stabilize
the magnetic state. 
To clarify the
internal structure of the spin-selective Kondo insulator, we firstly
look at static quantities.

As mentioned above, the ferromagnetic state in infinite dimensions 
has some intriguing
properties;\cite{peters2012} the one-particle spectral function of the
minority electrons shows a gap at the Fermi energy. The
mechanism of this gap formation is the commensurability arising from
the cooperation between localized- and conduction-electrons; 
they arrange themselves so that 
\begin{equation}
\langle n_{\downarrow,i}^f\rangle+\langle n_{\downarrow,i}^c\rangle=1
\end{equation}
holds for any lattice site $i$. Here we have assumed that 
 the localized moment is formed by a half filled {\it f}-orbital
and the spin polarization is written as
$\langle S_{z,i} \rangle=\frac{1}{2}(\langle n_{\uparrow,i}^f\rangle-\langle
n_{\downarrow,i}^f\rangle )$, which is the case for heavy fermion
materials. 

The essence of the commensurability may be understood clearly 
via the following arguments. Imagine first the Kondo 
lattice model at half filling, which shows the Kondo-gap formation 
under the commensurability condition: the sum of the averaged number
 of conduction and localized electrons at each site is unity per spin
 direction.  
This is the well-known mechanism of generating the Kondo insulator. 
For a system away from half filling, a correlated metallic 
phase is stabilized. However, if the system is very close to half 
filling, small magnetic fields can induce a spin polarization 
and simultaneously drive the system into a half-metallic (majority-spin 
conduction electrons) and half-insulating (minority-spin conduction 
electrons) state, where only the electrons in the insulating sector 
satisfies the commensurability (half filling) condition. This 
exemplifies the commensurability away from half filling in finite 
magnetic fields. What is nontrivial and remarkable for the spin-selective 
Kondo insulator is that the commensurability is generated concomitantly 
with the ferromagnetic transition. In particular, it emerges for a 
metallic system  far away from half filling, where there is no sign 
of the gap-formation in the normal metallic phase. Even in such 
low-filling cases, once the ferromagnetism is spontaneously induced 
by the Kondo exchange interaction, the emergent commensurability 
plays an essential role in generating a finite gap at the 
Fermi level, which in turn stabilizes the ferromagnetic state 
energetically.

For the 1D Kondo lattice model it has been known from
the studies for the one-electron case\cite{sigrist1991} and the exact
diagonalization calculations\cite{tsunetsugu1993}
that the quantum number of the ferromagnetic ground state for
strong enough coupling strength $J$ is
\begin{equation}
\mathcal{S}^{tot}=1/2\vert L-N\vert,\label{eqsum1}
\end{equation}
 where
$\mathcal{S}^{tot}$ denotes the total SU(2)-spin quantum number, $L$ the
number of  lattice sites, and $N$ the number of conduction electrons. 
In a classical picture, a state with this spin quantum number describes
the situation in which all localized spins are polarized in one
direction and all electrons point to the opposite direction. 
However, this is not the case for our model where the Kondo effect does work, reducing the polarization of conduction as well as localized electrons
via large quantum fluctuations.

We now show that the ground state having a spin
quantum number $\mathcal{S}^{tot}=1/2\vert L-N\vert$ in 1D
is a direct consequence of the emergent commensurability, which was 
introduced in the DMFT calculations.\cite{peters2012} 
Suppose that the symmetry is spontaneously broken 
and the maximal $S_z^{tot}$ value represents the homogeneous ground state for a 
given SU(2) spin quantum number $\mathcal{S}^{tot}$. 
Then we can derive the commensurability by just inserting the definitions
of the local quantities:
\begin{eqnarray*}
S_z^{tot}&=&1/2(N_c-N)\\
(N^f_\uparrow-N^f_\downarrow)+(N^c_\uparrow-N^c_\downarrow)&=&N^c-N\\
(n^f_\uparrow-n^f_\downarrow)+(n^c_\uparrow-n^c_\downarrow)&=&n^c_\uparrow+n^c_\downarrow-1\\
(n^f_\uparrow-n^f_\downarrow)&=&2n^c_\downarrow-1\\
2(n^c_\downarrow+n^f_\downarrow)&=&2.
\end{eqnarray*}
In the second line, the local expectation value is calculated by averaging
over all lattice sites (corresponding to a homogeneous
state). Furthermore a half-filled {\it f}-orbital is assumed.

We thus come to the conclusion: the fact that the ferromagnetic 
ground state in the 1D Kondo lattice model is associated with the
spin quantum number $\mathcal{S}^{tot}=1/2(L-N)$ is equivalent to the
emergent commensurability in our DMFT calculations.\cite{peters2012}
However, if one chooses a different state,
$\vert S_z^{tot}\vert<\mathcal{S}^{tot}$, as the ground state, the
commensurability is not easily visible, because the spin channels are mixed.
Therefore, if the commensurability condition is satisfied, the 
highest $S_z^{tot}$-value of the energetically degenerate states 
in Fig. \ref{DMRG_energ}, for which $\vert S_z^{tot}\vert
=\mathcal{S}^{tot}$ holds, should feature a straight line, $S_z^{tot}=1/2(L-N)$. 
One can see that this is indeed the case in Fig. \ref{DMRG_energ}.
Thus, the emergent commensurability allows us to use 
the relation $\mathcal{S}^{tot}=1/2(L-N)$ not only for the strong 
coupling regime as originally suggested,\cite{sigrist1991} but also for 
the weak coupling regime. 

\begin{figure}[tb]
\includegraphics[width=1\linewidth]{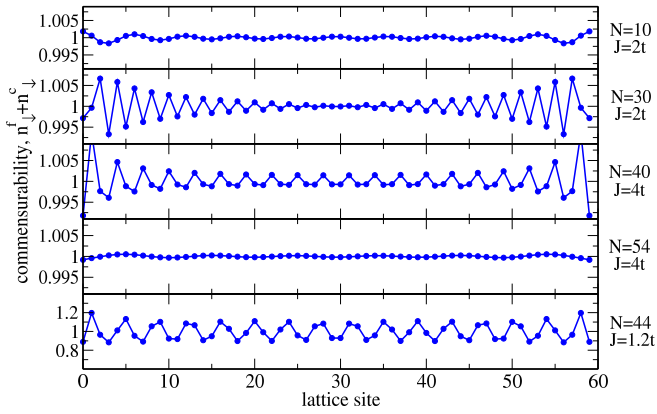}
\caption{(Color online) Locally measured commensurability, $\langle
  n^c_{\downarrow,i}+n^f_{\downarrow,i}\rangle=\langle
  n^c_{\downarrow,i}\rangle-\langle S_{z,i}\rangle+0.5$, in the
  finite size 1D
  Kondo lattice model. Calculations were performed for a $60$-sites chain using
  $m=800$ states. While the upper four panels show examples of the
  strong coupling ferromagnet, the lowest panel shows the ``second''
  ferromagnetic phase for intermediate couplings. The interaction
  parameters are given on the right side of each panel. Lines are
  guide to eye.
\label{oneD_commens}}
\end{figure}

We have performed calculations for a finite-size Kondo lattice
model. The introduction of open boundaries induces
oscillations in all measured quantities.
In Fig. \ref{oneD_commens} we show how the local commensurability
behaves depending on the lattice site in the ferromagnetic state with 
$S_z^{tot}=1/2(L-N)$, for an $L=60$-sites 1D Kondo lattice model.
Although the commensurability must be fulfilled when averaged over
all sites, it is not necessarily  fulfilled locally in a finite size
system with open boundaries. 
In Fig. \ref{oneD_commens} we can see that, for the
ferromagnetic states at strong coupling (upper $4$ panels), the
commensurability is almost fulfilled locally within one percent, where 
 small oscillations are due to open boundaries. On the other hand, 
the ``second''
ferromagnetic phase (lower panel) at intermediate coupling strengths
shows very strong 
oscillations. The overall tendency implies that oscillations are stronger
near quarter filling and weak coupling.

\begin{figure}[tb]
\includegraphics[width=0.49\linewidth]{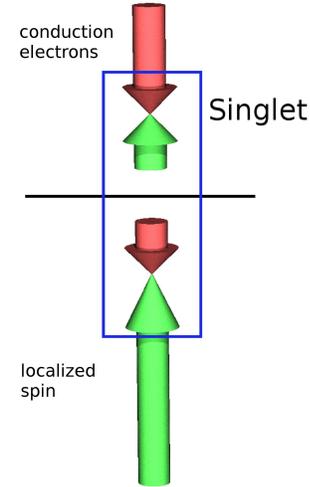}
\caption{(Color online) Sketch of a state fulfilling the
  commensurability. The upper 
  (lower) part corresponds to 
  the conduction electrons (localized spins). The commensurability is
  equivalent to the fact that there are the same number of
 minority conduction electrons
  (here: spin 
  up) and minority {\it f}-electrons
  (here spin down).
 \label{state1}}
\end{figure}

We present a schematic picture in Fig. \ref{state1}, which
may help to imagine how the ferromagnetic state looks under the 
emergent commensurability. Given that the localized spin is 
formed by a half-filled  strongly correlated {\it f}-electron, 
then the commensurability 
condition leads to the situation in which the number of 
minority conduction electrons (spin up in Fig. \ref{state1}) 
is equal to that of minority {\it f}-electrons (spin down in
Fig. \ref{state1}). Thus, the total number of spin up electrons becomes unity.
In this situation, the minority conduction electrons collaborate with the
minority localized electrons, together with the same fraction of opposite 
majority electrons, to form a Kondo singlet state, while the rest of
the majority conduction electrons and localized spins form a
ferromagnetic state.

\begin{figure}[tb]
\includegraphics[width=1\linewidth]{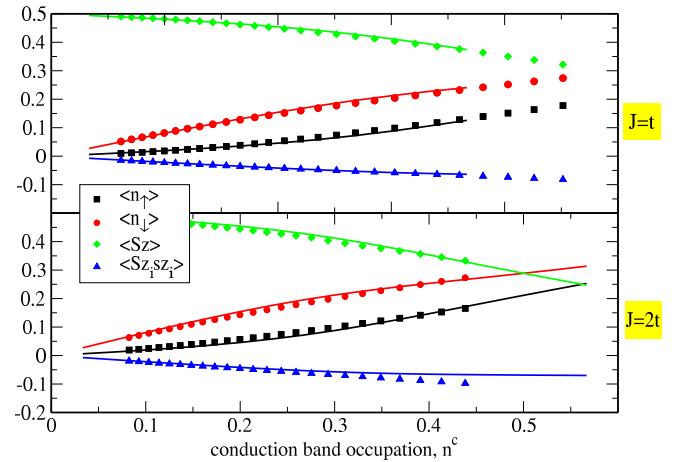}
\caption{(Color online) Local expectation values
  for different lattice fillings, averaged local occupation $n^c=\langle n^c_\uparrow\rangle
  +\langle n^c_\downarrow\rangle$ (1D: $n_c=N/L$ N: electron number L:
  length of the chain), in the ferromagnetic state with
  $S_z^{tot}=1/2(L-N)$. Continuous lines correspond to the 1D Kondo
  lattice model calculated for $L=60$-sites chains and averaged
  over all lattice sites. For comparison we include the 
  results of the previous
  DMFT calculations which are plotted as four different symbols
  (coupling strengths: upper panels $J=t$, lower panels  $J=2t$). 
 \label{comp_exps}}
\end{figure}

\begin{figure}[tb]
\includegraphics[width=1\linewidth]{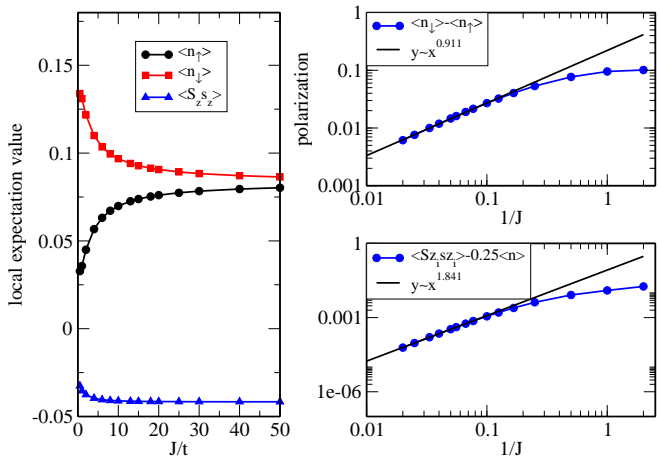}
\caption{(Color online) Local expectation values for the 1D Kondo lattice model, calculated for $L=60$ sites chain
  and $N=10$ electrons, thus $n=N/L\approx 0.17$ ($m=800$ states kept
  in DMRG). Right upper panel shows a power-law-fitting to the electron
  magnetization. The right lower panel shows a power-law-fitting to
  the spin-spin correlation function which approaches $-1/4\langle
  n_\uparrow+n_\downarrow\rangle$ in the strong coupling limit.
 (Lines are guide to the eye.)
 \label{oneD_bigj}}
\end{figure}

A close examination of local expectation values enables us to
 analyze the ferromagnetic state in more detail.
In Fig. \ref{comp_exps} and Fig. \ref{oneD_bigj} we compare
static expectation values  
for different fillings and different coupling strengths. The
expectation values are calculated 
for finite-size chains and averaged over all lattice sites.
The commensurability is fulfilled for all shown data points. Thus, the
expectation value of the localized spin $\la S_z\ra$ is directly related to
the occupation number of the minority conduction electrons.
The  expectation value $\la S_zs_z\ra$, which represents the
local spin-spin correlation between conduction electron and
localized spin, fulfills the condition $\langle S_z
s_z\rangle > -0.25\langle n_\uparrow+n_\downarrow\rangle$.
If all conduction electrons would be bound in local singlets this
spin-spin correlation is supposed to be equal to one quarter of the
occupation. If we associate these expectation values with the state
shown in Fig. \ref{state1}, then the polarization, $\la
n_\uparrow-n_\downarrow \ra$, corresponds to the number of 
majority electrons that are not bound in local singlets. Therefore, for
interaction strengths $\vert J\vert \approx \vert t\vert$ the number of electrons bound in 
local singlets and the number of polarized electrons are roughly equal.

In Fig. \ref{comp_exps} we also include the expectation values
obtained 
by DMFT calculations,\cite{peters2012} and find that
 the values of DMFT and DMRG are very close to each other.
This is remarkable because the phase boundaries 
calculated by DMFT show completely different behavior from the
present 1D case. Therefore, we can say that 
local expectation values of the ground state are not sensitive to the
dimensionality, and are controlled mainly by electron filling and
interaction strength. On the other hand, the transition point between
the paramagnet 
 and ferromagnet strongly depends on the lattice geometry.
This suggests that the contribution of the kinetic energy plays a key role in
determining which has the lowest energy,
  the ferromagnetic state or the paramagnetic state.

Let us now discuss how the ferromagnetic state behaves 
when increasing the interaction strength from weak to strong
coupling. In the strong coupling limit, the ferromagnetic phase in the
1D Kondo lattice model covers the whole phase diagram
except half filling where the system is paramagnetic.
In Fig. \ref{oneD_bigj} we compare local
expectation values in the 1D lattice for different
coupling strengths. The filling is fixed at $\la n\ra\approx 0.17$. 
It is seen that the electron polarization is larger for weak
coupling. With increasing coupling strength the polarization of the
electrons decreases and vanishes as a power law, as shown in the upper
right panel of Fig. \ref{oneD_bigj}. The exponent depends on the
electron filling.
In a similar way, the expectation value for the localized 
spin polarization (not shown), $\la S_z\ra$,
approaches a finite value, $\la S_z\ra =0.5-\la n_\uparrow\ra$,
 given by the commensurability. Also the local spin-spin correlation
$\la S_z s_z\ra$ between
the conduction electron and the spin approaches a fixed value, given
by $\la S_z s_z\ra \rightarrow -0.25(\la n_\uparrow\ra +\la
n_\downarrow\ra )$. All these quantities show power-law 
behavior in the strong coupling regime. The monotonic power-law decrease 
in the electron polarization, as well 
as $\la S_z s_z\ra \rightarrow -0.25(\la n_\uparrow\ra 
+\la n_\downarrow\ra )$, implies that most of electrons are bound 
into local singlets for large $J$.
In the limit of infinitely strong coupling $J$ where $\la
n_\downarrow\ra=\la n_\uparrow\ra$ and $\la S_z s_z\ra = -0.25(\la
n_\uparrow\ra +\la n_\downarrow\ra )$ all the 
conduction electrons are bound in local singlets.

\subsection{Spectral properties}

Having studied the static quantities, let us now look into
 spectral functions with particular focus on the gap 
created for the minority conduction electrons.
We have so far clarified that the 
mechanism of the gap formation in the ferromagnetic state 
is the emergent commensurability for the
minority conduction electrons. 
A glance at the energy values presented in
Fig. \ref{DMRG_energ} clearly shows that there is a jump in the energy between
the ferromagnetic state and the next excited state. Assuming that the
ground state of the 
system is the state with quantum number $S_z^{tot}=1/2(L-N)$ (in the
ferromagnetic phase), which is realized if an arbitrarily small
magnetic field is applied in the {\it z}-direction, then the single
particle spectral function for the conduction electrons is given by
excitations between the ferromagnetic ground state and excited states
with a quantum number $S_z^{tot}=1/2(L-N)+1$. In Fig. \ref{DMRG_energ},
there is an energy gap between these two states.
Furthermore, we can see that the second ferromagnetic phase at 
intermediate couplings also fulfills the commensurability, and
that there is an energy gap between the ground state and these
excited states.
\begin{figure}[tb]
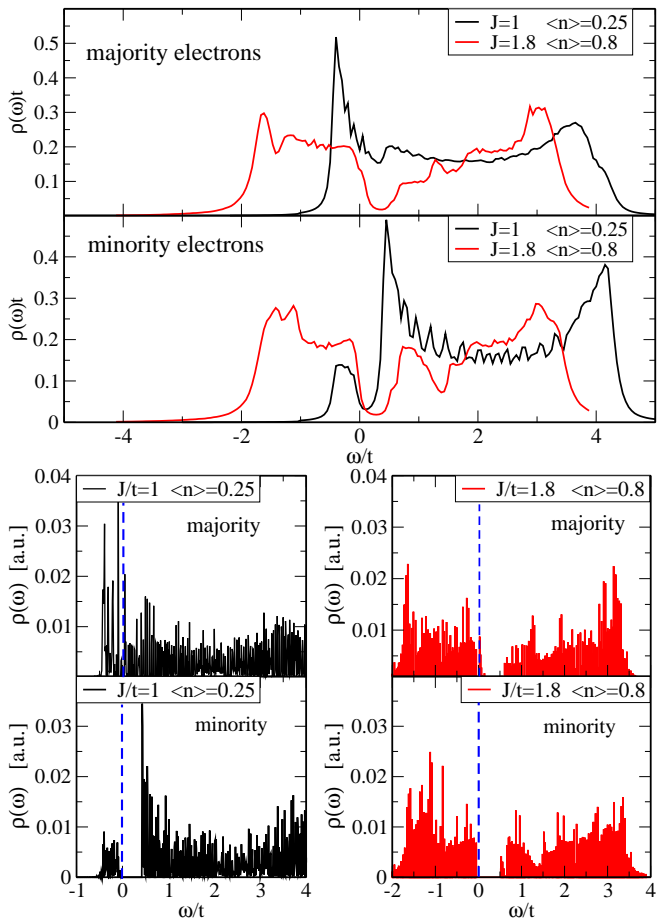

\includegraphics[width=\linewidth]{fig7a.eps}\\
\includegraphics[width=\linewidth]{fig7b.eps}
\caption{(Color online) Local spectral functions for
 the ferromagnetic state with $S_z=1/2(L-N)$ in the 1D Kondo lattice model. 
The spectral functions are calculated by the
correction vector method for an $L=40$-sites chain using $m=400$ states
and $\eta/t=0.05$. The Fermi energy lies at $\omega=0$.
Upper panels: Local spectral functions for ($J/t=1$, $\la n\ra=0.25$)
corresponding to a ferromagnetic state at low filling, and for
($J/t=1.8$,  $\la n\ra=0.8$) corresponding to a state in the second
ferromagnetic phase).
Lower panel:  Deconvoluted spectral functions (see text), which
describe the excitations in the spectral function. 
The blue line corresponds to the Fermi energy.
 \label{fig_Spec_loc}}
\end{figure}
\begin{figure}[tb]
\includegraphics[width=\linewidth]{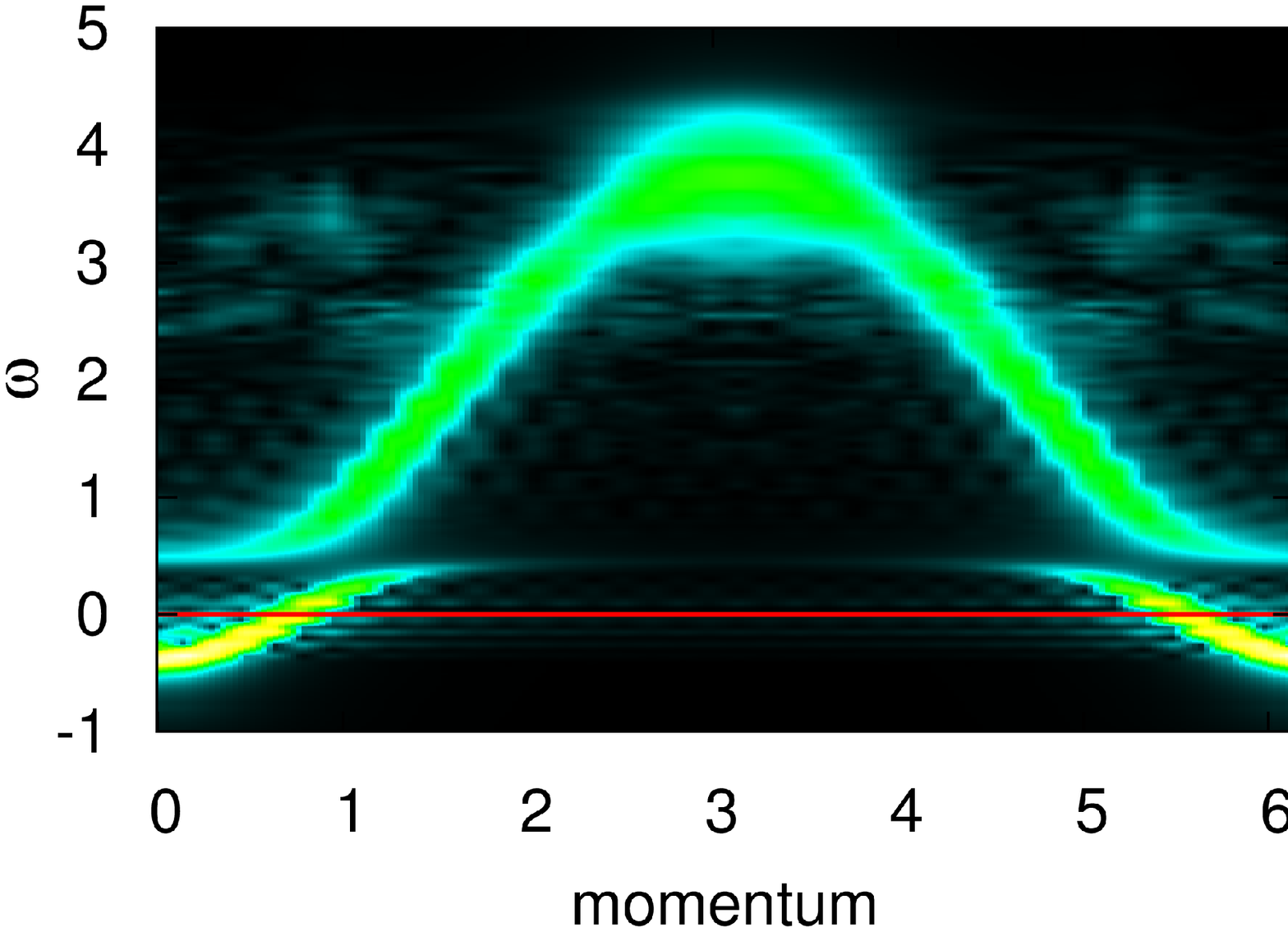}
\includegraphics[width=\linewidth]{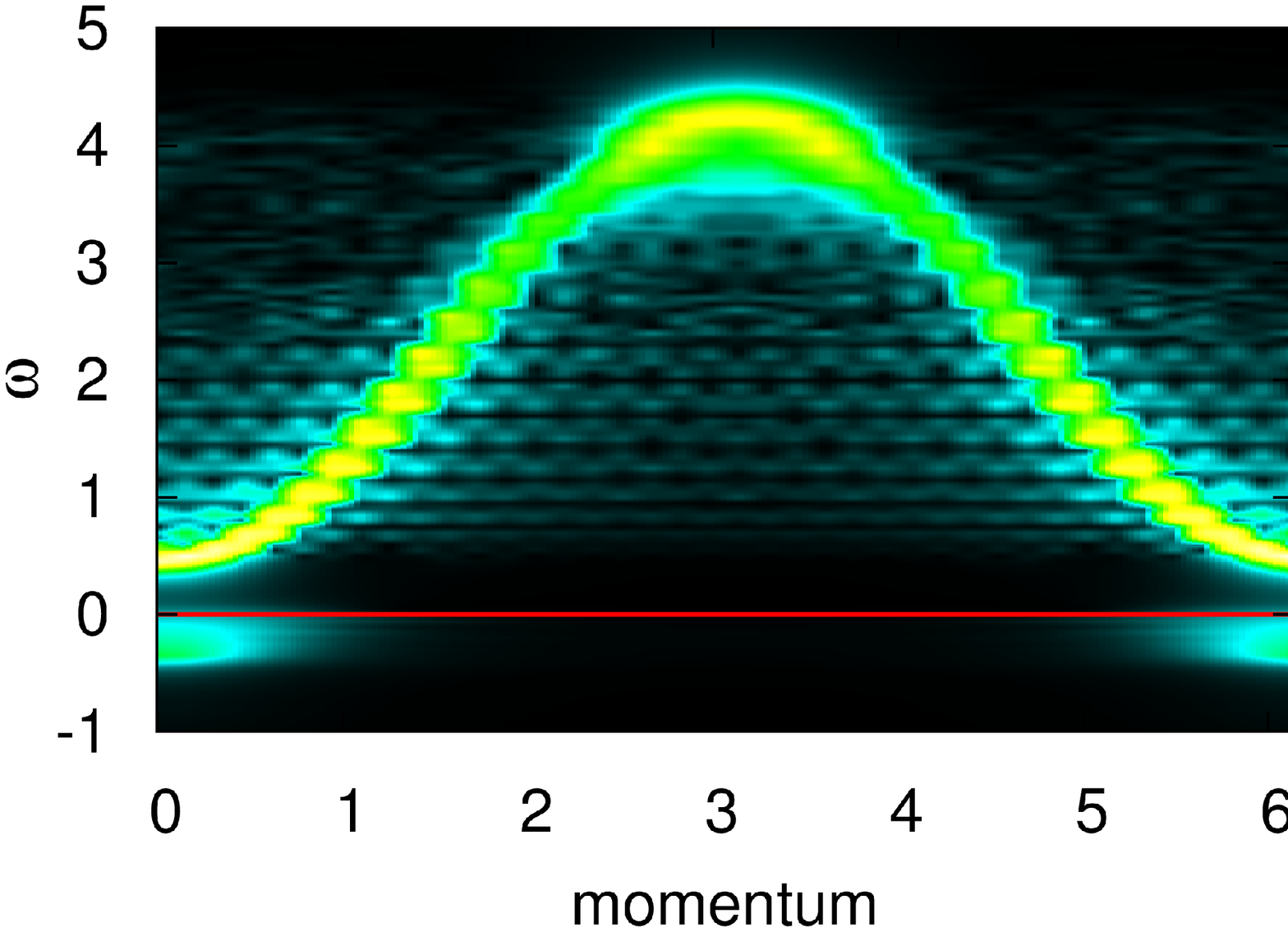}
\caption{(Color online) Momentum-resolved spectral functions for the
  1D Kondo lattice model in the ferromagnetic state. The
  red line represents the Fermi energy. The upper and lower panel
  show the spectral functions for the majority- and minority-conduction
  electrons, respectively. 
The interaction parameter and
  filling are  $J=t$ and $\la
  n\ra=0.25$. The spectral functions are calculated by the correction
  vector method for an $L=40$-sites chain using $m=400$ states and $\eta/t=0.05$.
 \label{fig_Spec_k}}
\end{figure}

We show in Fig. \ref{fig_Spec_loc} the local spectral functions of
conduction electrons in both
ferromagnetic phases. The state at ($J/t=1$, $\la n\ra=0.25$)
corresponds to the ferromagnetic state extending from low fillings,
while the state at ($J/t=1.8$, $\la n\ra=0.8$) corresponds to the
second ferromagnetic phase at intermediate coupling strengths.
For the calculation we have used the state
corresponding to $S_z^{tot}=1/2(L-N)$. 
We have performed a correction-vector
calculation\cite{kuhner1999,jeckelmann2002} with Lorentzian broadening
$\eta/t=0.05$. Because of this broadening the spectral
functions of the minority electrons (upper panels) show only a dip but
no real gap at the Fermi energy. To
improve the resolution we have calculated a delta-peak excitation spectrum
corresponding to a deconvoluted spectral function.\cite{peters_spec2011}
It is now seen from the computed excitation spectrum 
that the spectral functions of the minority conduction electrons indeed have a
gap at the Fermi energy for both ferromagnetic phases in 
the 1D Kondo lattice model. Note that 
the ferromagnetic state at intermediate coupling strengths 
also has a gap for the majority conduction electrons, but
slightly above the Fermi energy. This is the expected behavior, as
both spin channels should be gapped for the half-filled lattice. 
Although the mechanism for the formation of two ferromagnetic phases
might be different, the ground state shows very similar 
properties fulfilling the commensurability and having a gap for the
minority conduction electrons.

In Fig. \ref{fig_Spec_k}, we show the momentum
resolved spectral function for ($J/t=1$, $\la n\ra=0.25$). 
The shape of the spectrum looks very similar for both spin directions. However,
while the lower band for the minority conduction electrons terminates below
the Fermi energy,  the lower band for the majority conduction electrons
crosses the Fermi energy. To see the gap in the spectrum, it is
important to choose $S_z^{tot}=\pm 1/2(L-N)$ as a quantum number for the
ground state. Note that 
in the previous DMRG calculations with SU(2) symmetry,\cite{smerat2009}
the gap at the Fermi energy is not visible, because both spin
channels are mixed. 

\begin{figure}[tb]
\includegraphics[width=\linewidth]{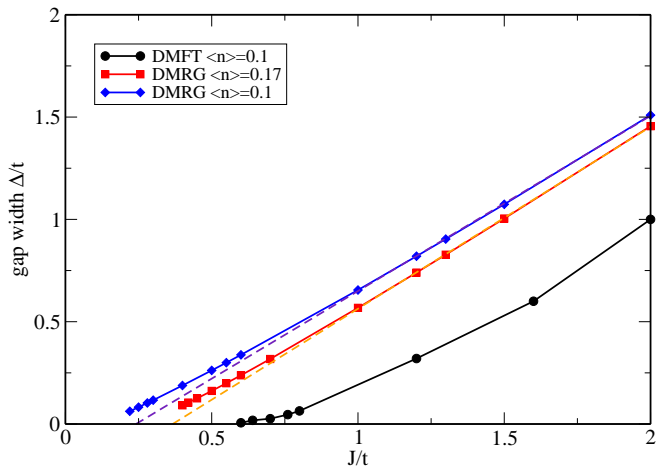}
\caption{(Color online) Comparison of the gap width between the
  ferromagnetic state in infinite dimensions (DMFT results) and one
  dimension (DMRG results). The dashed lines represent linear fits to the
  strong coupling region. 
 \label{gap}}
\end{figure}

Finally, in Fig. \ref{gap}
we compare the gap width for different interaction strengths.
We also include the results from our DMFT calculations.\cite{peters2012}
It is seen that the gap in the DMFT results nicely follows a
Kondo-type functional form $T_K\sim \exp(-1/J)$. 
 The curves for the 1D Kondo lattice model shown in
 Fig. \ref{gap}, which may not follow an exponential-type
behavior, terminate at the phase 
boundary to the paramagnetic solution at weak coupling.
Generally, the gap width of the 1D Kondo lattice model is
larger than in the DMFT results.
Furthermore, the gap width in
the 1D lattice is linear in the coupling strength for
strong coupling, showing derivations in the vicinity of the phase
boundary. 
As these derivations occur only in a very small region around
the phase boundary, it is very hard to tell the functional form of the
gap width in this region.
However, the deviation from an exponential form can be understood 
from the known fact that the Kondo temperature in a  1D Tomonaga-Luttinger
liquid takes a power-law form depending on the system
parameters.\cite{furusaki1994} We therefore believe that the gap width
corresponds to the energy scale of the Kondo temperature in a 1D lattice.

\section{Summary}

We have studied the ferromagnetic state in the 1D Kondo
lattice model. Using the density matrix renormalization group, we have
calculated the phase diagram and confirmed the existence of 
two ferromagnetic phases. Motivated by our recent findings 
based on DMFT in infinite dimensions, we have
focused on the spectral functions in the 
ferromagnetic state. It has been elucidated that, similar to DMFT, 
the spin selective Kondo insulator 
is indeed realized in the 1D lattice, in which the minority
conduction electrons form a Kondo gap at the Fermi energy, whereas the
majority conduction electrons stay metallic. This gap is stabilized by 
the emergent commensurability:
the number of minority conduction electrons, together with the localized 
electrons with the same spin, should be unity in the ferromagnetic phase.
It has also been shown that the commensurability is equivalent to the
statement that the ground state in the ferromagnetic phase is given by
a state with spin quantum number $\mathcal{S}^{tot}=1/2(L-N)$. 
The commensurability and the corresponding gap formation 
have indeed been observed in the two distinct ferromagnetic
phases of the 1D Kondo lattice model. 
Through these comprehensive analyses in one and infinite dimensions,
 we have confirmed  that  the spin-selective Kondo insulator is 
the fundamental notion for the ferromagnetic state 
of the Kondo lattice model in any dimension.

\begin{acknowledgments}
We acknowledge fruitful discussions with A. Koga and Y. Tada,
T. Pruschke, G. Khaliullin, S. Hoshino,
Y. Kuramoto, and  N. Shibata.
RP thanks the Japan Society for the Promotion of Science (JSPS)
and the Alexander von Humboldt-Foundation for support during a 2 years
fellowship and for the following support from JSPS by its FIRST Program.
NK is
supported by KAKENHI (No. 20102008) and JSPS through its
FIRST Program.  The calculations were performed at the ISSP in Tokyo.
\end{acknowledgments}


\end{document}